\documentclass[12pt]{article}
 \usepackage{epsfig}
 \def\be{\begin{equation}}
 \def\ee{\end{equation}}
 \def\bea{\begin{eqnarray}}
 \def\eea{\end{eqnarray}}
 \usepackage{graphicx}

 \catcode`\@=11
 \def\lsim{\mathrel{\mathpalette\@versim<}}
 \def\gsim{\mathrel{\mathpalette\@versim>}}
 \def\@versim#1#2{\vcenter{\offinterlineskip
 \ialign{$\m@th#1\hfil##\hfil$\crcr#2\crcr\sim\crcr } }}
 \catcode`\@=12

 \catcode`@=12
\begin{document}
 \thispagestyle{empty}
 \begin{flushright}
 UCRHEP-T573\\
 November 2016\
 \end{flushright}
 \vspace{0.6in}
 \begin{center}
 {\LARGE \bf Generalized Gauge U(1) Family Symmetry\\
 for Quarks and Leptons\\}
 \vspace{1.2in}
 {\bf Corey Kownacki, Ernest Ma, Nicholas Pollard, and Mohammadreza Zakeri\\}
 \vspace{0.2in}
 {\sl Physics and Astronomy Department,\\ 
 University of California, Riverside, California 92521, USA\\}
 \end{center}
 \vspace{1.2in}

\begin{abstract}\
If the standard model of quarks and leptons is extended to include three 
singlet right-handed neutrinos, then the resulting fermion structure 
admits an infinite number of anomaly-free solutions with just one simple 
constraint.  Well-known examples satisfying this constraint are $B-L$, 
$L_\mu - L_\tau$, $B-3L_\tau$, etc.   We derive this simple constraint, 
and discuss two new examples which offer some insights to the structure of 
mixing among quark and lepton families, together with their possible 
verification at the Large Hadron Collider.
\end{abstract}

 \newpage
 \baselineskip 24pt

\noindent \underline{\it Introduction}~:~\\
In the standard model of particle interactions, there are three families 
of quarks and leptons.  Under its $SU(3)_C \times SU(2)_L \times U(1)_Y$ 
gauge symmetry, singlet right-handed neutrinos $\nu_R$ do not transform.  
They were thus not included in the minimal standard model which only has 
three massless left-handed neutrinos.  Since neutrinos are now known to be 
massive, $\nu_R$ should be considered as additions to the standard model.  
In that case, the 
model admits a possible new family gauge symmetry $U(1)_F$, with charges 
$n_{1,2,3}$ for the quarks and $n'_{1,2,3}$ for the leptons as shown in 
Table 1.
\begin{table}[htb]
\caption{Fermion assignments under $U(1)_F$.}
\begin{center}
\begin{tabular}{|c|c|c|c|c|}
\hline
Particle & $SU(3)_C$ & $SU(2)_L$ & $U(1)_Y$ & $U(1)_F$ \\
\hline
$Q_{iL} = (u,d)_{iL}$ & 3 & 2 & 1/6 & $n_i$ \\
$u_{iR}$ & $3$ & 1 & $2/3$ & $n_i$ \\
$d_{iR}$ & $3$ & 1 & $-1/3$ & $n_i$ \\
\hline
$L_{iL} = (\nu,l)_{iL}$ & 1 & 2 & $-1/2$ & $n'_i$ \\
$l_{iR}$ & 1 & 1 & $-1$ & $n'_i$ \\
$\nu_{iR}$ & 1 & 1 & 0 & $n'_i$ \\
\hline
\end{tabular}
\end{center}
\end{table}

To constrain $n_{1,2,3}$ and $n'_{1,2,3}$, the requirement of gauge anomaly 
cancellation is imposed.  The contributions of color triplets to the 
$[SU(3)]^2 U(1)_F$ anomaly sum up to
\begin{equation}
[SU(3)]^2 U(1)_F~: ~~~ {1 \over 2} \sum_{i=1}^3 (2n_i - n_i - n_i);
\end{equation}
and the contributions of $Q_{iL},u_{iR},d_{iR},L_{iL},l_{iR}$ to the 
$U(1)_Y [U(1)_F]^2$ anomaly sum up to
\begin{equation}
U(1)_Y [U(1)_F]^2~: ~~~ \sum^3_{i=1} \left[ 6 \left( {1 \over 6} \right) - 
3 \left( {2 \over 3} \right) - 3 \left( -{1 \over 3} \right) \right] n_i^2 
+ \left[ 2 \left( -{1 \over 2} \right) - (-1) \right] {n'_i}^2.
\end{equation}
Both are automatically zero, as well as the $[U(1)_F]^3$ anomaly because 
all fermions couple to $U(1)_F$ vectorially.  The contributions of the 
$SU(2)_L$ doublets to the $[SU(2)]^2 U(1)_F$ anomaly sum up to
\begin{equation}
[SU(2)]^2 U(1)_F~: ~~~ {1 \over 2} \sum^3_{i=1} (3n_i + n'_i);
\end{equation}
and the contributions to the $[U(1)_Y]^2 U(1)_F$ anomaly sum up to
\begin{eqnarray}
[U(1)_Y]^2 U(1)_F &:&  \sum^3_{i=1} \left[ 6 \left( {1 \over 6} \right)^2 - 
3 \left( {2 \over 3} \right)^2 - 3 \left( -{1 \over 3} \right)^2 \right] n_i 
+ \left[ 2 \left( -{1 \over 2} \right)^2 - (-1)^2 \right] n'_i \nonumber \\ 
&=& \sum^3_{i=1} \left( -{3 \over 2} n_i - {1 \over 2} n'_i \right).
\end{eqnarray}
Both are zero if
\begin{equation}
\sum^3_{i=1} (3n_i + n'_i) = 0.
\end{equation}

There are many specific examples of models which satisfy this condition as 
shown in Table 2.
\begin{table}[htb]
\caption{Examples of models satisfying Eq.~(5).}
\begin{center}
\begin{tabular}{|c|c|c|c|c|c|c|}
\hline
$n_1$ & $n_2$ & $n_3$ & $n'_1$ & $n'_2$ & $n'_3$ & Model \\
\hline
1/3 & 1/3 & 1/3 & $-1$ & $-1$ & $-1$ & $B-L$~\cite{mm80} \\
\hline
0 & 0 & 0 & 0 & 1 & $-1$ & $L_\mu-L_\tau$~\cite{hjlv91,mrr02,agpy14,hhrs15} \\
\hline
1/3 & 1/3 & 1/3 & 0 & 0 & $-3$ & $B-3L_\tau$~\cite{m98,ms98,mr98,bhmz10} \\
\hline
1/3 & 1/3 & 1/3 & 3 & $-3$ & $-3$ & Ref.~\cite{lm10} \\
\hline
1 & 1 & $-2$ & 1 & 1 & $-2$ & Ref.~\cite{ltw12} \\
\hline
$a$ & $a$ & $-2a$ & 0 & $-1$ & 1 & Ref.~\cite{cdh15} \\
\hline
\end{tabular}
\end{center}
\end{table}
If there are four families, then $n_{1,2,3}=1/3$, $n_4=-1$, and 
$n'_{1,2,3}=1$, $n'_4=-3$, would also satisfy Eq.~(5).  This may then be 
considered~\cite{fpw10} as the separate gauging of $B$ and $L$.

In this paper, we discuss two new examples which offer some insights to the 
structure of mixing among quarks and lepton families.  Both have nontrivial 
connections between quarks and leptons.  Their structures are shown in Table 3.
\begin{table}[htb]
\caption{Two new models satisfying Eq.~(5).}
\begin{center}
\begin{tabular}{|c|c|c|c|c|c|c|}
\hline
$n_1$ & $n_2$ & $n_3$ & $n'_1$ & $n'_2$ & $n'_3$ & Model \\
\hline
1 & 1 & 0 & $0$ & $-2$ & $-4$ & A \\
\hline
1 & 1 & $-1$ & $0$ & $-1$ & $-2$ & B \\
\hline
\end{tabular}
\end{center}
\end{table}
In both cases, with only one Higgs doublet with zero charge under $U(1)_F$, 
quark and lepton mass matrices are diagonal except for the first two quark 
families.  This allows for mixing among them, but not with the third family. 
It is a good approximation to the $3 \times 3$ quark mixing matrix, to the 
extent that mixing with the third family is known to be suppressed. 
In the lepton sector, mixing also comes from the Majorana mass matrix 
of $\nu_R$ which depends on the choice of singlets with vacuum expectation 
values which break $U(1)_F$.  Adding a second Higgs doublet with nonzero 
$U(1)_F$ charge will allow 
mixing of the first two families of quarks with the third in both cases. 
As for the leptons, this will not affect Model A, but will cause mixing 
in the charged-lepton and Dirac neutrino mass matrices in Model B.  
Flavor-changing neutral currents are 
predicted, with interesting phenomenological consequences.

\noindent \underline{\it Basic structure of Model A}~:~\\
Consider first the structure of the $3 \times 3$ quark mass matrix 
${\cal M}_d$ linking $(\bar{d}_L,\bar{s}_L,\bar{b}_L)$ to $(d_R,s_R,b_R)$.  
Using
\begin{equation}
\Phi_1 = (\phi_1^+,\phi_1^0) \sim (1,2,1/2;0),
\end{equation}
with $\langle \phi_1^0 \rangle = v_1$, it is clear that ${\cal M}_d$  is 
block diagonal with a $2 \times 2$ submatrix which may be rotated on the 
left to become
\begin{equation}
{\cal M}_d = \pmatrix{c_L & -s_L & 0 \cr s_L & c_L & 0 \cr 0 & 0 & 1} 
\pmatrix{m'_d & 0 & 0 \cr 0 & m'_s & 0 \cr 0 & 0 & m'_b},
\end{equation}
where $s_L = \sin \theta_L$ and $c_L = \cos \theta_L$. 
We now add a second Higgs doublet
\begin{equation}
\Phi_2 = (\phi_2^+,\phi_2^0) \sim (1,2,1/2;1),
\end{equation}
with $\langle \phi_2^0 \rangle = v_2$, so that
\begin{equation}
{\cal M}_d = \pmatrix{c_L & -s_L & 0 \cr s_L & c_L & 0 \cr 0 & 0 & 1} 
\pmatrix{m'_d & 0 & m'_{db} \cr 0 & m'_s & m'_{sb} \cr 0 & 0 & m'_b}
\end{equation}
is obtained.
At the same time, ${\cal M}_u$ is of the form
\begin{equation}
{\cal M}_u = \pmatrix{m'_u & 0 & 0 \cr 0 & m'_c & 0 \cr m'_{ut} & m'_{ct} & m'_t} 
\pmatrix{c_R & s_R & 0 \cr -s_R & c_R & 0 \cr 0 & 0 & 1},
\end{equation}
where it has been rotated on the right.  Because of the physical mass hierarchy 
$m_u << m_c << m_t$, the diagonalization of Eq.~(10) will have very small 
deviations from unity on the left.  Hence the unitary matrix diagonalizing 
Eq.~(9) on the left will be essentially the experimentally observed 
quark mixing matrix $V_{CKM}$ which has three angles and one phase. 
Now ${\cal M}_d$ of Eq.~(9) has exactly seven parameters, the 
three diagonal masses $m'_d, m'_s, m'_b$, the angle $\theta_L$, the 
off-diagonal mass $m'_{sb}$ which can be chosen real, and the off-diagonal mass 
$m'_{db}$ which is complex.  With the input of the three quark mass 
eigenvalues $m_d, m_s, m_b$ and $V_{CKM}$, these seven parameters can be 
determined.  
 
Consider the diagonalization of the real mass matrix
\begin{equation}
\pmatrix{a & 0 & s_1 c \cr 0 & b & s_2 c \cr 0 & 0 & c} = V_L 
\pmatrix{ a(1-s_1^2/2) & 0 & 0 \cr 0 & b(1-s_2^2/2) & 0 \cr 0 & 0 & 
c(1+s_1^2/2+s_2^2/2)} V_R^\dagger,
\end{equation}
where $s_{1,2} << 1$ and $a << b << c$ have been assumed.  We obtain
\begin{equation}
V_L = \pmatrix{1-s_1^2/2 & -s_1 s_2 b^2/(b^2-s_1^2c^2-a^2) & s_1 \cr 
s_1 s_2 a^2/(b^2+s_2^2c^2-a^2) & 1-s_2^2/2 & s_2 \cr -s_1 & -s_2 & 
1-s_1^2/2-s_2^2/2},
\end{equation}
and
\begin{equation}
V_R^\dagger = \pmatrix{1 & s_1 s_2 ab/(b^2-a^2) & -s_1 a/c \cr 
-s_1 s_2 ab/(b^2-a^2) & 1 & -s_2 b/c \cr s_1 a/c & s_2 b/c & 1}.
\end{equation}
Hence
\begin{equation}
V_{CKM} = \pmatrix{c_L & -s_L & 0 \cr s_L & c_L & 0 \cr 0 & 0 & 1} 
\pmatrix{e^{i \alpha} & 0 & 0 \cr 0 & 1 & 0 \cr 0 & 0 & 1} V_L,
\end{equation}
where $\alpha$ is the phase transferred from $m'_{db}$.

Comparing the above with 
the known values of $V_{CKM}$~\cite{pdg16}, we obtain
\begin{equation}
s_1 = 0.00886, ~~~ s_2 = 0.0405, ~~~ s_L = -0.2253, ~~~ e^{i \alpha} = 
-0.9215 + i 0.3884,
\end{equation}
with $m_d = m'_d$, $m_s = m'_s$, $m_b = m'_b$ to a very good approximation.

\noindent \underline{\it Scalar sector of Model A}~:\\ 
In addition to $\Phi_{1,2}$, we add a scalar singlet
\begin{equation}
\sigma \sim (1,1,0;1),
\end{equation}
then the Higgs potential containing $\Phi_{1,2}$ and $\sigma$ is given by
\begin{eqnarray}
V &=& m_1^2 \Phi_1^\dagger \Phi_1 + m_2^2 \Phi_2^\dagger \Phi_2 + m_3^2 
\bar{\sigma} \sigma + [\mu \sigma \Phi_2^\dagger \Phi_1 + H.c.] \nonumber \\ 
&+& {1 \over 2} \lambda_1 (\Phi_1^\dagger \Phi_1)^2 + 
{1 \over 2} \lambda_2 (\Phi_2^\dagger \Phi_2)^2 + 
{1 \over 2} \lambda_3 (\bar{\sigma} \sigma)^2 + 
\lambda_{12} (\Phi_1^\dagger \Phi_1)(\Phi_2^\dagger \Phi_2) \nonumber \\ 
&+& \lambda'_{12} (\Phi_1^\dagger \Phi_2)(\Phi_2^\dagger \Phi_1) + 
\lambda_{13} (\Phi_1^\dagger \Phi_1)(\bar{\sigma} \sigma) + 
\lambda_{23} (\Phi_2^\dagger \Phi_2)(\bar{\sigma} \sigma). 
\end{eqnarray}
Let $\langle \phi^0_{1,2} \rangle = v_{1,2}$ and $\langle \sigma \rangle = u$, 
then the minimum of $V$ is determined by
\begin{eqnarray}
0 &=& v_1 ( m_1^2 + \lambda_1 v_1^2 + (\lambda_{12} + \lambda'_{12})v_2^2 + 
\lambda_{13} u^2) + \mu v_2 u, \\
0 &=& v_2 ( m_2^2 + \lambda_2 v_2^2 + (\lambda_{12} + \lambda'_{12})v_1^2 + 
\lambda_{23} u^2) + \mu v_1 u, \\
0 &=& u ( m_3^2 + \lambda_3 u^2 + \lambda_{13} v_1^2 + 
\lambda_{23} v_2^2) + \mu v_1 v_2.
\end{eqnarray}
For $m_2^2$ large and positive, a solution exists with $v_2^2 << v_1^2 << u^2$, 
i.e.
\begin{eqnarray}
u^2 \simeq {-m_3^2 \over \lambda_3}, ~~~ v_1^2 \simeq {-m_1^2 - 
\lambda_{13} u^2 \over \lambda_1}, ~~~ v_2 \simeq {-\mu v_1 u \over m_2^2 + 
\lambda_{23} u^2}.
\end{eqnarray}
Hence the scalar particle spectrum of Model A consists of a Higgs boson $h$ 
very much like that of the SM with $m_h^2 \simeq 2 \lambda_1 v_1^2$, a 
heavy Higgs boson which breaks $U(1)_F$ with $m_\sigma^2 \simeq 2 \lambda_3 u^2$, 
and a heavy scalar doublet very much like $\Phi_2$ with 
$m^2 (\phi_2^+,\phi_2^0) \simeq m_2^2 + \lambda_{23} u^2$.

\noindent \underline{\it Gauge sector of Model A}~:\\ 
With the scalar structure already considered, the $Z-Z_F$ mass-squared 
matrix is given by
\begin{equation}
{\cal M}^2_{Z,Z_F} = \pmatrix {g_Z^2 (v_1^2 + v_2^2)/4 & -g_Z g_F v_2^2/2 \cr 
-g_Z g_F v_2^2/2 & g_F^2 (u^2 + v_2^2)}.
\end{equation}
The $Z-Z_F$ mixing is then $(g_Z/2g_F)(v_2^2/u^2)$.  For $v_2 \sim 10$ GeV 
and $u \sim 1$ TeV, this is about $10^{-4}$, well within the experimentally 
allowed range.

Since $Z_F$ couples to quarks and leptons according to $n_{1,2,3}$ and 
$n'_{1,2,3}$, its branching fractions to $e^-e^+$ and $\mu^- \mu^+$ are 
given by $2{n'_{1,2}}^2/(12\sum n_i^2 + 3\sum {n'_i}^2)$.  Since $n'_1 = 0$, 
we need consider only the branching fraction $Z_F \to \mu^- \mu^+$ 
to compare against data.  For Model A, it is about 2/21. 
The $c_{u,d}$ coefficients used in the experimental 
search~\cite{atlas14_z,cms15_z} of $Z_F$ are then
\begin{equation}
c_u = c_d = 2g_F^2 (2/21).
\end{equation}
For $g_F = 0.13$, a lower bound of about 4.0 TeV on $m_{Z_F}$ is obtained 
from the Large Hadron Collider (LHC) based on the preliminary 13 TeV data 
by comparison with the published data from the 7 and 8 TeV runs.  
Note however that if $Z_F \to e^- e^+$ is ever observed, this particular 
model is ruled out.

\noindent \underline{\it Flavor-changing interactions}~:\\
Whereas the SM $Z$ boson does not mediate any flavor-changing interactions, 
the heavy $Z_F$ does because it distinguishes families.  For quarks,
\begin{equation}
{\cal L}_{Z_F} = g_F Z_F^\mu (\bar{u}' \gamma_\mu u' + \bar{c}' \gamma_\mu c' 
 + \bar{d}' \gamma_\mu d' + \bar{s}' \gamma_\mu s' ). 
\end{equation}
Using Eqs.~(12) and (13) to express the above in terms of mass eigenstates 
for the $d$ sector, and keeping only the leading flavor-changing terms, 
we find
\begin{equation}
{\cal L}'_{Z_F} = g_F Z_F^\mu [s_1 (\bar{d}_L \gamma_\mu b_L + \bar{b}_L 
\gamma_\mu d_L) + s_2 (\bar{s}_L \gamma_\mu b_L + \bar{b}_L 
\gamma_\mu s_L) - s_1 s_2 (\bar{d}_L \gamma_\mu s_L + \bar{s}_L 
\gamma_\mu d_L)]. 
\end{equation}
From the experimental values of the $B^0-\bar{B}^0$, $B_S^0-\bar{B}_S^0$, 
and $K_L-K_S$ mass differences, severe constraints on $g_F^2/m_{Z_F}^2$ 
are obtained, coming from the operators
\begin{equation}
(\bar{d}_L \gamma_\mu b_L)^2 + H.c., ~~~ (\bar{s}_L \gamma_\mu b_L)^2 + H.c., 
~~~ (\bar{d}_L \gamma_\mu s_L)^2 + H.c. 
\end{equation}
respectively.  Using typical values of quark masses and hadronic decay and 
bag parameters~\cite{flav10}, we estimate the various Wilson coefficients 
to find their contributions as follows:
\begin{eqnarray}
\Delta M_{B} &=& 4.5 \times 10^{-2} ~s_1^2 (g_F^2/m^2_{Z_F})~{\rm GeV}^3, \\ 
\Delta M_{B_s} &=& 6.4 \times 10^{-2} ~s_2^2 (g_F^2/m^2_{Z_F})~{\rm GeV}^3, \\ 
\Delta M_K &=& 1.9 \times 10^{-3} ~s_1^2 s_2^2 (g_F^2/m^2_{Z_F})~{\rm GeV}^3.
\end{eqnarray}
Using Eq.~(15) and assuming that the above contributions are no more than 
10\% of their experimental values~\cite{pdg16}, we find the lower limits 
on $m_{Z_F}/g_F$ to be 10.2, 9.5, 0.84 TeV respectively.  This is easily 
satisfied for $m_{Z_F} > 4.0$ TeV with $g_F = 0.13$ from the LHC bound discussed 
in the previous section.

In the scalar sector, since $\Phi_{1,2}$ both contribute to ${\cal M}_d$, 
the neutral scalar field orthogonal to the SM Higgs field will also 
mediate flavor-changing interactions.  The Yukawa interactions are
\begin{equation}
{\cal L}_Y = {h_1 \over \sqrt{2} v_1} (m'_d \bar{d}'_L d'_R + 
m'_s \bar{s}'_L s'_R + m'_b \bar{b}'_L b'_R) +  {h_2 \over \sqrt{2} v_2} 
(m'_{db} \bar{d}'_L b'_R + m'_{sb} \bar{s}'_L b'_R). 
\end{equation} 
Extracting again the leading flavor-changing terms, we obtain
\begin{eqnarray}
{\cal L}'_Y &=& \left( {h_2 \over \sqrt{2} v_2} - {h_1 \over \sqrt{2} v_1} 
\right) (s_1 m_b \bar{d}_L b_R + s_2 m_b \bar{s}_L b_R - s_1 s_2 m_s  
\bar{d}_L s_R - s_1 s_2 m_d \bar{s}_L d_R \nonumber \\ 
&& ~~~~~~~~~~~~~~~~~~~~ -s_1 s_2^2 m_d \bar{b}_L d_R - s_2^3 m_s \bar{b}_L s_R), 
\end{eqnarray}
where the physical scalar $(v_1h_2 - v_2h_1)/\sqrt{v_1^2 + v_2^2} = H + iA$ 
is a complex field, with $m_H \simeq m_A$.

Assuming negligible mixing between $H$ or $A$ with the SM $h$ (identified 
as the 125 GeV particle observed at the LHC), we consider the following 
effective operators~\cite{mm13}:
\begin{eqnarray}
&& {s_1^2 m_b^2 \over 8 v_2^2} \left( {1 \over m_H^2} - {1 \over m_A^2} 
\right) (\bar{d}_L b_R)^2 - {s_1^2 s_2^2 m_b m_d \over 4 v_2^2} 
\left( {1 \over m_H^2} + {1 \over m_A^2} \right) (\bar{d}_L b_R) 
(\bar{d_R} b_L) + H.c., \\ 
&& {s_2^2 m_b^2 \over 8 v_2^2} \left( {1 \over m_H^2} - {1 \over m_A^2} 
\right) (\bar{s}_L b_R)^2  - {s_2^4 m_b m_s \over 4 v_2^2} 
\left( {1 \over m_H^2} + {1 \over m_A^2} \right) (\bar{s}_L b_R) 
(\bar{s_R} b_L) + H.c., \\ 
&& {s_1^2 s_2^2 m_s^2 \over 8 v_2^2} \left( {1 \over m_H^2} - {1 \over m_A^2} 
\right) (\bar{d}_L s_R)^2  - {s_1^2 s_2^2 m_s m_d \over 4 v_2^2} 
\left( {1 \over m_H^2} + {1 \over m_A^2} \right) (\bar{d}_L s_R) 
(\bar{d_R} s_L) + H.c. 
\end{eqnarray}
The upper bounds on $(1/v_2^2)[(1/m_H^2)-(1/m_A^2)]$ from 
$\Delta M_B, \Delta M_{B_s}, \Delta M_K$ are then
\begin{equation}
(4.5 \times 10^{-9}, ~ 5.3 \times 10^{-9}, ~ 4.5 \times 
10^{-3})~{\rm GeV}^{-4},
\end{equation}  
respectively, whereas those on $(1/v_2^2)[(1/m_H^2)+(1/m_A^2)]$ are 
\begin{equation}
(1.4 \times 10^{-4}, ~ 1.7 \times 10^{-5}, ~ 8.0 \times 
10^{-5})~{\rm GeV}^{-4}.
\end{equation} 
For $v_2 = 10$ GeV, these are easily satisfied with 
for example $m_H = 500$ GeV and $m_A = 520$ GeV.

\noindent \underline{\it Lepton sector of Model A}~:~
With the chosen $U(1)_F$ charges $(0,-2,-4)$ of Table 3, the charged-lepton 
and Dirac neutrino mass matrices (${\cal M}_l$ and ${\cal M}_D$) are both 
diagonal.  As for the $3 \times 3$ Majorana mass matrix ${\cal M}_R$ of 
$\nu_R$, it depends on the choice of scalar singlets which break $U(1)_F$.  
We have already used $\sigma \sim 1$ [see Eq.~(16)] to induce a small 
$v_2$ [see Eq.~(21)].  Call that $\sigma_1$ and add $\sigma_{2,4} \sim 2,4$, 
with vacuum expectation values $u_{1,2,4}$ respectively. Then 
\begin{equation}
{\cal M}_R = \pmatrix{M_0 & M_1 & M_2 \cr M_1 & M_3 & 0 \cr M_2 & 0 & 0},
\end{equation}
where $M_0$ is an allowed invariant mass term, $M_1$ comes from $u_2$, 
and $M_{2,3}$ from $u_4$.  The seesaw neutrino mass matrix is then
\begin{equation}
{\cal M}_\nu = {\cal M}_D {\cal M}_R^{-1} {\cal M}_D^T = \pmatrix{0 & 0 & a 
\cr 0 & b & c \cr a & c & d},
\end{equation}
where the two texture zeros appear because of the form of ${\cal M}_R$ 
and ${\cal M}_D$ being diagonal~\cite{m05}.  This form is known to 
be suitable for a best fit~\cite{lmw14} to current neutrino-oscillation data 
with normal ordering of neutrino masses.

\noindent \underline{\it Basic structure of Model B}~:~
The quark structure of Model B is basically the same as that of Model A, 
with the second Higgs doublet now having two units of $U(1)_F$ charge, i.e.
\begin{equation}
\Phi_2 = (\phi_2^+,\phi_2^0) \sim (1,2,1/2;2).
\end{equation}
Hence $\sigma_2 \sim (1,1,0;2)$ is needed for the 
$\sigma_2 \Phi_2^\dagger \Phi_1$ term in Eq.~(17).

In the gauge sector, again $Z_F \to e^-e^+$ is zero, and the branching fraction 
$Z_F \to \mu^-\mu^+$ is now 2/51.  The $c_{u,d}$ coefficients are then
\begin{equation}
c_u = c_d = 2 g_F^2(2/51).
\end{equation}
For the same choice of $g_F = 0.13$ for Model A, the present experimental 
lower bound from LHC data is reduced from 4.0 TeV to 3.7 TeV.  For quarks,
\begin{equation}
{\cal L}_{Z_F} = g_F Z_F^\mu (\bar{u}' \gamma_\mu u' + \bar{c}' \gamma_\mu c' 
 - \bar{t}' \gamma_\mu t' + \bar{d}' \gamma_\mu d' + \bar{s}' \gamma_\mu s' 
 - \bar{b}' \gamma_\mu b'). 
\end{equation}
Using Eqs.~(12) and (13) to express the above in terms of mass eigenstates 
for the $d$ sector, and keeping only the leading flavor-changing terms, 
we find
\begin{equation}
{\cal L}'_{Z_F} = 2g_F Z_F^\mu [-s_1 (\bar{d}_L \gamma_\mu b_L + \bar{b}_L 
\gamma_\mu d_L) - s_2 (\bar{s}_L \gamma_\mu b_L + \bar{b}_L 
\gamma_\mu s_L) + s_1 s_2 (\bar{d}_L \gamma_\mu s_L + \bar{s}_L 
\gamma_\mu d_L)]. 
\end{equation}
This differs from Eq.~(25) only by an overall factor of $-2$. 
As for the scalar sector, Eqs.~(30) and (31) remain the same.

\noindent \underline{\it Lepton sector of Model B}~:~
With the chosen $U(1)_F$ charges $(0,-1,-2)$ of Table 3, the charged-lepton 
and Dirac neutrino mass matrices are given by
\begin{equation}
{\cal M}_l = \pmatrix{m'_e & 0 & m'_{e \tau} \cr 0 & m_\mu & 0 \cr 0 & 0 & 
m'_\tau}, ~~~ 
{\cal M}_D = \pmatrix{m'_1 & 0 & 0 \cr 0 & m'_2 & 0 \cr m'_{31} & 0 & 
m'_3}. 
\end{equation}
Using the scalar singlets $\sigma_1 \sim 1$ as well $\sigma_2$, the $\nu_R$ 
Majorana mass matrix is again given by Eq.~(37).  Now even though ${\cal M}_D$ 
is not diagonal, Eq.~(38) is still obtained, thereby guaranteeing a best fit 
to current neutrino-oscillation data.  The difference from Model A is the 
presence of $\tau-e$ transitions from the nondiagonal ${\cal M}_l$. 
However, for $m'_{e \tau}/m'_\tau < 0.1$, the branching fraction 
of $\tau \to e \mu^- \mu^+$ is less than $2 \times 10^{-11}$, far below 
the current bound of $4.1 \times 10^{-8}$.

\noindent \underline{\it Application to LHC anomalies}~:~
Whereas $Z_F$ also mediates $b \to s \mu^- \mu^+$, its effect is too small 
in Models A and B to explain the tentative LHC observations of $B \to K^* 
\mu^- \mu^+$ and the ratio of $B^+ \to K^+ \mu^- \mu^+$ to 
$B^+ \to K^+ e^- e^+$~\cite{hmn16}. 
The reason is the stringent bound on $m_{Z_F}$ from LHC data as a function 
of $g_F$ through the parameters $c_{u,d}$ of Eqs.~(23) and (40).  Suppose 
we take $n_{1,2,3} = (0,0,1)$ and $n'_{1,2,3} = (0,-3,0)$, then $Z_F$ 
couples to 
only $\mu^- \mu^+$ and $b' \bar{b}'$, thus allowing for $b-s$ mixing, 
but $c_{u,d}=0$.  This evades the direct LHC bound, and may be used to 
explain the $B$ anomalies if they persist.  Of course, Eqs.~(27) to (29) 
still hold, and a full analysis of the detailed structure of $B \to K^* 
\mu^- \mu^+$ will be required.

\noindent \underline{\it Conclusion}~:~
We have generalized the $B-L$ symmetry as a gauge $U(1)_F$ extension of the 
standard model, where quarks and leptons of each family may transform 
differently.  We have considered two new examples (A and B), each with two 
Higgs doublets and restricted quark mass matrices consistent with data. 
The new $Z_F$ gauge boson couples differently to each quark and lepton 
family, and is constrained by present data to be heavier than about 4 TeV 
if $g_F = 0.13$.  Future data may reveal just such a $Z_F$ belonging to 
this class of models.  Flavor-changing interactions are suitably 
suppressed by the assignments of quarks and leptons under $U(1)_F$. 
In the leptonic sector, with the addition of a minimal set of Higgs 
singlets, a Majorana neutrino mass matrix of two texture zeros may 
be obtained, leading to a best fit of neutrino-oscillation data with 
normal ordering of neutrino masses.  

\noindent \underline{\it Acknowledgement}~:~
This work was supported in part by the U.~S.~Department of Energy Grant 
No. DE-SC0008541.

\bibliographystyle{unsrt}

\end{document}